\documentclass[12pt]{article}
\usepackage{rotating}
\usepackage{epsfig,amssymb}
\input{epsf}

\def\laq{\raise 0.4 ex \hbox{$<$}\kern -0.8 em\lower 0.62 ex\hbox{$\sim$}}
\def\gaq{\raise 0.4 ex \hbox{$>$}\kern -0.7 em\lower 0.62 ex\hbox{$\sim$}}

\def\beq{\begin{equation}}
\def\eeq{\end{equation}}
\def\beqa{\begin{eqnarray}}
\def\eeqa{\end{eqnarray}}

\def\to{\rightarrow}

\def\to{\rightarrow}

 \def\frac#1#2{{\textstyle{{#1}\over {#2}}}}
 \def\lsim{\mathrel{\rlap{\lower4pt\hbox{\hskip1pt$\sim$}}
    \raise1pt\hbox{$<$}}} \def\gsim{\mathrel{\rlap{\lower4pt\hbox{\hskip1pt$\sim$}}
    \raise1pt\hbox{$>$}}}
\def\sqr#1#2{{\vcenter{\vbox{\hrule height.#2pt
         \hbox{\vrule width.#2pt height#1pt \kern#1pt
         \vrule width.#2pt}
         \hrule height.#2pt}}}}

\def\gappeq{\mathrel{\rlap {\raise.5ex\hbox{$>$}} {\lower.5ex\hbox{$\sim$}}}}

\def\lappeq{\mathrel{\rlap{\raise.5ex\hbox{$<$}}
{\lower.5ex\hbox{$\sim$}}}}

\begin{document}
\pagestyle{plain}

\begin{flushright}
October 2021
\end{flushright}
\vspace{15mm}

\begin{center}

{\Large\bf Inflation, phase transitions and the cosmological constant}

\vspace*{1.0cm}

Orfeu Bertolami$^*$\\
\vspace*{0.5cm}
{Departamento de F\'\i sica e Astronomia, Faculdade de Ci\^encias, Universidade do Porto\\
Rua do Campo Alegre, 4169-007 Porto, Portugal}\\

\vspace*{2.0cm}
\end{center}

\begin{abstract}

\noindent

Cosmological phase transitions are thought to have taken place at the early Universe imprinting their properties on the observable Universe. There is strong evidence that, through the dynamics of a scalar field that  lead a second order phase transition, inflation shaped the Universe accounting for the most conspicuous features of the observed Universe. It is argued that inflation has also striking implications for the vacuum energy. Considerations for subsequent second order phase transitions are also discussed.

\end{abstract}

\vfill
\noindent\underline{\hskip 140pt}\\[4pt]
\noindent
{$^{*}$ Also at Centro de F\'\i sica das Universidades do Minho e do Porto, Rua do Campo Alegre, 4169-007 Porto, Portugal} \\
{E-mail address: orfeu.bertolami@fc.up.pt}

\newpage
\section{Introduction}
\label{sec:intro}

Inflation endows the Universe with its most salient observational features. One of the first realisations of inflation involved a first order cosmological phase transition driven by the Higgs field of a  Grand Unified Theory (GUT) where bubbles of the new vacuum nucleated at the old vacuum of the GUT. The completion of this scenario required a successful expansion of the new vacuum bubbles, their collision and an efficient percolation so to homogeneously fill and replace entirely the old vacuum \cite{Guth}. However, it was soon understood that inflation driven by the Higgs field of a GUT could not warrant all these conditions and, furthermore, it did not allow for quantum field fluctuations to generate the needed energy density fluctuations to form structures. Therefore, models with a potential that allowed for a second order phase transition were since then favoured \cite{Albrecht,Linde} (see e.g. Ref. \cite{Olive} for a review). After inflation the Universe is supercooled and a suitable mechanism for its reheating must follow in order to lead to a hot Universe compatible with nucleosynthesis, the cosmic microwave background radiation and the observed large scale structure of the Universe.   

In what concerns the vacuum energy, as pointed out by Zeldovich \cite{Zeldovich} long ago, it gravitates and from Lorentz invariance it implies that it must be included into Einstein's field equations as cosmological constant, that is, its energy-momentum tensor has the form $T_{ab} = {\Lambda \over 8 \pi G} g_{ab}$. Thus, the associated vacuum energy, $\rho_{V}$, corresponds to the $T_{00}$ component while the vacuum isotropic pressure, $p_V$, to the $T_{ii}$ components. Explicit computations of the zero-point energy in the momentum space in the context of field theory (see, for instance, Ref. \cite{Akhmedov}) imply that these components must satisfy the equation of state:
\beq
p_{V} = - \rho_{V}~~.
\label{eq:eqstate}
\eeq
Field theory estimates lead to the well known conclusion that a severe fine tuning is needed into the Einstein's field equations to match the observed value: $\Omega_{\Lambda} \simeq 0.7$, or
$\rho_{V} \simeq 5.6 \times 10^{-47}~GeV^4$ for $h_0=0.7$, which compares with
the contribution from the Standard Model (SM) after the Higgs field acquired a non-vanishing vacuum
expectation value, $\rho_{V}^{(SM)} = {\cal O}(250~GeV)^4$, a discrepancy of a factor $10^{56}$ \cite{Dreitlein,Linde2,Veltman}. GUTs and quantum gravity lead to even more dramatic discrepancies. This is the well known cosmological constant problem (see, for instance, Refs. \cite{Weinberg,Carroll,Sahni-Starobinsky,Bertolami09} for reviews and discussions).  For sure, this is a problem that arises due to the non-trivial quantum nature of the vacuum. 
However, this fundamental quantum origin manifests itself macroscopically through its equation of state (\ref{eq:eqstate}) and we argue that it is affected by inflation subjected to thermodynamical considerations.  As we shall see these considerations suggest that the vacuum must have a passive temperature which increases during inflation and hence the vacuum can be seen as a heat sink. 
We point out that attempts to "sequester" the vacuum energy through a volumetric dilution were previously considered, but only for a closed space-time \cite{Kaloper,Bertolami20}. 

In what follows we shall argue that generic thermodynamic considerations for the vacuum together with the well known Bekenstein bound for entropy \cite{Bekenstein} in the context of the new inflationary scenario imply in a significant suppression of the vacuum energy once the inflationary phase transition is completed. This is a crucial point as the vacuum entropy is often assumed to vanish. Further arguments will be presented to support that similar considerations might apply to subsequent second order cosmological phase transitions.

\section{The inflationary phase transition} \label{sec:inflationary phase transition}

Let us now focus on the transition that took place during the inflationary process where, following Coleman and co-workers \cite{Coleman}, we distinguish the old or ``false" vacuum energy before inflation, $\rho_F$, that is dynamically driven into the new ``true" vacuum energy, $\rho_T$ ,after inflation. The matter sector of the Universe was then dominated by a generic inflaton field, initially at the top of its potential at a thermal bath at temperature $T$. 

The starting point to describe the vacuum evolution process is the Gibbs-Duhem equation:
 \beq
S dT - V dp + \sum_{j=1}^n N_j d\mu_j = 0~~,
\label{eq:GDequ}
\eeq
where, $S$, $N_j$ and $\mu_j$ ($j=1, ..., n$) are the entropy, the number of particles and the chemical potential of each species, respectively . These quantities refer to the vacuum, but in order not to overburden the notation no explicit notation will be introduced to distinguish their nature but the difference between the "true" and "false" vacuum energy densities. As the two vacua, presumably have no free particles, $N_j=0$, from which follows that:
\beq
{dp \over dT} = {S \over V}~~.
\label{eq:GDVequ}
\eeq

In order to estimate the vacuum entropy we consider the Bekenstein entropy bound \cite{Bekenstein} as it was devised to hold for any physical system:
\beq
S  \leq {2 \pi k R E \over \hbar c}~~,
\label{eq:Bekenstein}
\eeq
where $k$ is Boltzmann's constant, $R$ is the system length scale and $E$ its energy. We saturate this bound considering for the length scale the horizon distance; hence Eq. (\ref{eq:GDVequ}) can be written as ($\hbar=c=k=1$):
\beq
{dp \over dT}  = {2 \pi  R_{Hor} E \over V}=  2 \pi  R_{Hor}  \rho~~.
\label{eq:BekensVequ}
\eeq
Assuming that inflation takes place due to a scalar field at an energy scale, $\Delta$, hence the expansion rate is given approximately by $H_I=\Delta^2/M$, where $M$ is the reduced Planck mass ($M=M_P/\sqrt{8\pi}$), and thus $a(t)=a_ie^{H_It}$ and $a_i$ is the scale factor at the onset of inflation. It follows that 
\beq
R_{Hor}= {1 \over H_I}[e^{-H_It_i} e^{H_It}-1], 
\eeq
where for an eternal De Sitter space, $t_i \to -\infty$. 

Therefore, using Eq. (\ref{eq:eqstate}), we get after integration: 
\beq
{p_T \over p_F} = e^{-2 \pi  I} ~~,
\label{eq:Thermodynr1}
\eeq
where in order to estimate $I$ we consider that the vacuum is an ordered state, whose temperature can either vanishing or to be very hot, that is, it tends to minus infinite. Thus, considering $T_i=0$,  assuming the physical limit, $T_f<-M$, and, consistently, the suitable integration measure, $dT=[\delta(t-t_i) T_i + \delta(t-t_f) (-T_f)]dt$, hence:
\beq
I={1 \over a_i H_I} \left[a_i T_i + a_f (-T_f)\right] > {a_f M  \over a_i H_I}~~,
\label{eq:I}
\eeq
where the index $f$ refers to the end of inflation. 

Therefore, writing the result in terms of the highest possible false vacuum energy, $\rho_F \simeq {\cal O}(M^4)$, then for $65$ e-foldings of inflation, $a_f=a_i e^{65}$,
\beq
\rho_T \simeq e^{(-2 \pi { e^{65} M \over  H_I})} M^4~~,
\label{eq:Thermodynr2}
\eeq
which is a substantial suppression as $H_I \simeq \Delta^2/M$, $\Delta$ being the inflaton potential scale, $V=\Delta^4$, and realistic inflation requires ${\cal O}(\Delta/M) \simeq 10^{-3}$. Indeed, it follows that the exponential suppression is significant, ${\cal O}(10^{-3 \times 10^{34}})$ .   

This result has at least three relevant implications. The obvious first one is that no fine tuning is now required. The second one it that dark energy, assumed to be responsible for the late time accelerated expansion of the Universe, is not due to any residual left by the quantum gravity vacuum or the GUT vacuum. The third consequence is that the vacuum is hotter that the matter content of the Universe whose temperature dropped dramatically during inflation. This requires that inflation is followed by a process of reheating driven by the coupling of the inflaton to the other fields. The mechanism here discussed suggests, in opposition, that the so-called warm inflationary models \cite{Berera} might be the most suitable one to fully describe the inflation scenario, which is in this instance no longer adiabatic.  This point will be further discussed in section 4.    

\section{Subsequent phase transitions} \label{sec:subsequent phase transitions}

The mechanism described above can account for the suppression of the cosmological constant after inflation, however the cosmological constant may pick other contributions, for instance, from the electroweak phase transition and the quark-gluon phase transition. It is of course impossible to apply the above considerations to subsequent phase transitions and furthermore, specific knowledge of these transitions should be known. However, some generic features of the mechanism outlined above can considered as a suitable guideline. In fact, an exponential change of the pressure arising from Eq. (\ref{eq:GDVequ}) can be found, for example, in the Clausius-Clapeyron equation for the change of the critical points with pressure and in the Kelvin-Helmholtz equation for the pressure change due to capillaries in fluids. In the  Clausius-Clapeyron equation, the argument of the exponential contains essentially the ratio between the latent heat to the thermal energy. In the Kelvin-Helmholtz equation, the argument of the exponential involves the ratio of the surface tension times the curvature of the meniscus to the thermal energy, and a pressure suppression arises when the meniscus is concave. A generic way to express the main features of these equations would be:
\beq
ln\left({p_T \over p_F}\right) = -{E_{eff} \over E_{Th}}~~,
\label{eq:Thermodynr3}
\eeq
where $E_{eff}$ is a generic energy difference that effectively characterise the phase transition and $E_{Th}$ is the thermal energy. A significant suppression might ensue whether $E_{eff}/E_{Th}  \gg 1$. If this behaviour can be encountered in cosmological phase transitions after inflation, a suppression of the vacuum energy may ensue.   

Unfortunately, it seems very difficult to advance with general conclusions without specific details of the phase transition in question. However, a conspicuous property of the phase transition implied by the new inflationary type models is that it involves a {\it spinodal decomposition}, in opposition to the nucleation process that characterises the first order phase transitions. The generic properties of the second order phase transitions are described by the Cahn-Hilliard model \cite{Cahn-Hilliard}. In this model the free energy is given in terms of the gradient of the concentration, $c(x)$, of each phase, where $x$ is a typical length scale between the phases. For a two domains system, the two phases correspond to $c=\pm 1$. The free energy variation is given by \cite{Cahn-Hilliard}: 
\beq
\Delta F= \int {\gamma \over 2} |\nabla c|^2 dV~~,
\label{eq:Cahn-Hilliard}
\eeq
where $\sqrt\gamma$ sets the length scale of the transition regions between the domains. Using  
\beq
p =-\left({\partial F \over \partial V}\right)_T ~~,
\label{eq:Cahn-Hilliard}
\eeq
in order to relate with the vacuum pressure and assuming the equilibrium solution of the diffusion equation of the model, the Cahn-Hilliard equation, $c(x) = tanh(x/\sqrt{2\gamma})$, hence for $x \gg 0$, $p \simeq 0$ ($p \to 0$ for $x \to \infty$) and therefore, the vacuum energy tends to vanish. 

Of course, these considerations are very general and may not hold under a more detailed analysis, however they do not seem to be ruled out neither by theoretical considerations \cite{Kibble,Linde3,Ginsparg} nor by some specific numerical simulations \cite{March-Russell,Nicolis,Ogilvie}. In what concerns the most relevant phase transition after inflation, the electroweak phase transition, it is well known that is it not sufficiently strong first order to support baryogenesis (see eg. Ref. \cite{Sami21} for an updated review and references therein).  In fact, baryogenesis is a quite relevant issue as it requires not only a specific creation mechanism, but also that it occurs after inflation and that any entropy generating process is prevented. It is worth mentioning that the thermodynamic suppression of anti-baryons can be achieved once the Sakharov conditions are meet, for instance, in the context of models where CPT symmetry is spontaneously violated \cite{Bertolami97}.

\section{Discussion and Conclusions}

The vacuum is an elusive concept. From just a featureless ground state in classical field theory and quantum mechanics it acquires a rich structure in quantum field theory. The pioneering efforts of Dirac, Zeldovich, Hawking, Coleman, Kibble and others have shown that the vacuum gravitates, it can give origin to a thermal radiation, undergo to phase transitions and can give rise to topological defects likewise in material systems. The quantum vacuum can also be at the core of the spontaneous breaking of symmetries, a crucial feature of the Standard Model of Fundamental Interactions of Nature. However, this prominent role poses an embarrassing problem of fine tuning, the cosmological constant problem, whose solution has been the object of countless attempts from a wide range of points of view. Given the need to bring gravity within the context of a quantum framework, it is often argued that the cosmological constant problem cannot be properly addressed without a suitable fundamental quantum gravity theory. It might be very well the case, however attempts in the context of, for instance, string theory, one of the most accomplished proposals to quantise gravity and to unify all interactions of Nature, has not shown to be successful in this respect \cite{Witten}, despite the interesting ideas that spring from the vacua landscape of string theory \cite{Susskind,Polchinski}, from M-theory compactifications \cite{BoussoPol} and other multiverse considerations \cite{Bertolami08}. The same can be stated about other proposals to quantise gravity such as loop quantum gravity, although recent hopeful and interesting results \cite{Zhang}.  

Our proposal, on the other hand, is much more modest. We argue that the universal laws of thermodynamics should also apply to the vacuum, as already considered for black holes and in attempts to endow the gravitational field with entropy \cite{Penrose}, and furthermore, that inflation provides a suitable way out for the fine tune problem of the the vacuum energy. Thus, as the inflaton evolves and drives the inflationary accelerated expansion that shapes most of the observable properties of the Universe, it also induces thermodynamical transformations in the vacuum that yield in its suppression. The scenario discussed in section 2 suggests that inflation is not an adiabatic process, which has common features with the warm inflation scenario, even though the latter does not provide the needed vacuum energy suppression. This suppression arises from the assumption that the vacuum itself is not a thermodynamically passive system and, in fact, the exponential suppression, Eq. (\ref{eq:Thermodynr2}), suggests that it has a multiply connected capillary like structure. This justifies the measure used to get the result Eq. (\ref{eq:I}). Indeed, the idea that the vacuum evolves \cite{Bronstein,Bertolami86,Ozer} in a smooth way with the cosmic time, $t$, actually as $\Lambda \sim t^{-2}$, has been shown to allow, in specific contexts, for interpolating the expected $\cal{O}$$(M_P^2)$ initial value of the cosmological constant to its observational value. Thus, if the vacuum evolves, it is not at all that surprising that an abrupt and significant event such as inflation can have an equally relevant impact on the vacuum energy. The arguments presented in section 2 seem to support this conjecture. It remains to be seen whether the presented assumptions can also be used for generic cosmological phase transitions, as proposed in section 3; if so, the several avatares of the cosmological constant can all be tackled with the same underlying set of thermodynamic arguments. 

Clearly, our proposal implies that the observed accelerated expansion of the Universe is not due to the cosmological constant and must be due to some field. There is no shortage of proposals: quintessence; k-essence; generalised Chaplygin gas, etc. (see Ref. \cite{Copeland} for a review). This has the observational signature that, at present epoch of dominance of dark energy, most likely $w \equiv {p \over \rho} \neq -1$. This implication could be actually become a direct connection if a single scalar field  would drive inflation and the late time accelerated expansion of the Universe, as in the so-called quintessential inflation (see e.g. \cite{PV,DV,BD}), even though specific proposals along these lines are still too involved and an underlying mechanism to connect these two widely different energy scales is still missing.  
 
In conclusion we should stress that the paradigmatic nature of inflation is reinforced. Indeed, besides the well known virtues of solving the initial conditions problems of the Big Bang model, of explaining the absence of topological defects such as magnetic monopoles and of giving rise to the energy density fluctuations responsible for structure formation, the arguments presented in this work indicate that the accelerated expansion provided by inflation induces changes in the vacuum that can suppress its energy density considerably. Thus, given that inflation is such a fundamental ingredient for solving so many issues, the question of the initial conditions that allow for its onset might hold the key to discriminate the viability scenarios arising from the very early Universe quantum gravity epoch.






\bibliographystyle{unstr}

\end{document}